\documentclass[12pt,a4paper]{article}

\usepackage[utf8]{inputenc}
\usepackage[english]{babel}
\usepackage{amsmath}
\usepackage{amsfonts} 
\usepackage{amssymb}
\usepackage{graphicx}
\usepackage{lmodern}
\usepackage{hyperref}
\usepackage[top=25mm,bottom=25mm]{geometry}

\title{Regional estimates of reproduction numbers with application to COVID-19
	\thanks{This work was funded by the Federal Ministry of Education and Research (BMBF; grant 05M18SIA).}} %

\author{%
	Jan Pablo Burgard%
	\thanks{Universität Trier, Germany
		(\texttt{burgardj@uni-trier.de}).
	}
	\and
	Stefan Heyder%
	\thanks{Technische Universit{\"{a}}t Ilmenau, Germany
		(\texttt{stefan.heyder@tu-ilmenau.de}, \texttt{thomas.hotz@tu-ilmenau.de}).
		}
	\and
	Thomas Hotz\footnotemark[3]
	\and
	Tyll Krueger%
    \thanks{Wroclaw Unversity of Science and Technology, Poland
        (\texttt{tyll.krueger@pwr.wroc.pl}).
    }
}

\begin{document}
	\maketitle
	
	\begin{abstract}
        In the last year many public health decisions were based on real-time monitoring the spread of the ongoing COVID-19 pandemic.
For this one often considers the reproduction number which measures the amount of secondary cases produced by a single infectious individual.
While estimates of this quantity are readily available on the national level, subnational estimates, e.g. on the county level, pose more difficulties since only few incidences occur there.
However, as countermeasures to the pandemic are usually enforced on the subnational level, such estimates are of great interest to assess the efficacy of the measures taken, and to guide future policy.\\
We present a novel extension of the well established estimator \cite{fraser_estimating_2007} of the country level reproduction number to the county level by applying techniques from small-area estimation.
This new estimator yields sensible estimates of reproduction numbers both on the country and county level. 
It can handle low and highly variable case counts on the county level, and may be used to distinguish local outbreaks from more widespread ones.
We demonstrate the capabilities of our novel estimator by a simulation study and by applying the estimator to German case data.

    \end{abstract}
	
\section{Introduction}
\label{sec:intro}
The ongoing COVID-19 pandemic is affecting countries worldwide with over 4.4 million deaths as of 30 August 2021 \cite{who_report_2021}. 
To restrict the spread of SARS-CoV-2, the virus causing COVID-19, many countries have implemented non-pharmaceutical countermeasures such as bans of mass gatherings, mandatory wearing of masks and reduction of contacts in the private and work life. 
In addition vaccines which reduce both the severity of COVID-19 and the infectiousness of vaccinated individuals have become available, and most European countries have vaccinated large portions of their population \cite{ecdc_vaccine_2021}.

To quantify the spread of an epidemic, one considers the time-varying reproduction number $ R(t) $, the mean amount of secondary cases a primary case infected on day $ t $ is expected to infect during his course of infection.
Knowing $ R(t) $ allows one to infer whether the number of cases will rise or fall in the future; the threshold for growth being $ R(t) = 1 $.
On the country-level a standard model for the spread of an epidemic is the following stochastic renewal equation for $ I(t) $, the amount of newly infected cases on day $ t $, which are assumed to be (conditionally) Poisson distributed:
\begin{equation}
    \label{eq:renewal}
    I(t) ~|~I(t - 1), \dots  ~\sim~ \mathrm{Pois} \left( R(t) \sum\limits_{\tau = 1}^\infty I(t - \tau) w(\tau) \right)\;.
\end{equation}
Here, $ w(\cdot) $ specifies the distribution of the generation time, i.e., given that a primary case infects a secondary case, $ w(\tau) $ is the probability that this infection occurs on day $ \tau $ after the primary case was infected himself.
A well studied estimator of $ R(t) $ in this model is 
\begin{equation}
    \label{eq:rhat}
   \hat R(t) = I(t) / \sum_{\tau = 1}^\infty I(t - \tau) w(\tau)\;,
\end{equation}
see e.g. \cite{fraser_estimating_2007,gostic_practical_2020}.
For this estimator to be reliable the denominator has to be large enough, as its variance (conditional on past cases) is $ R(t) / \sum_{\tau = 1}^\infty I(t-\tau) w(\tau)$, see \cite{hotz_monitoring_2020}.

A deficit of estimating the  reproduction number on the country level is that these estimates are affected by local outbreaks which, in the absence of high case numbers, dominate even country-level estimations.
\begin{figure}[t]
    \centering
    \includegraphics[width=\textwidth]{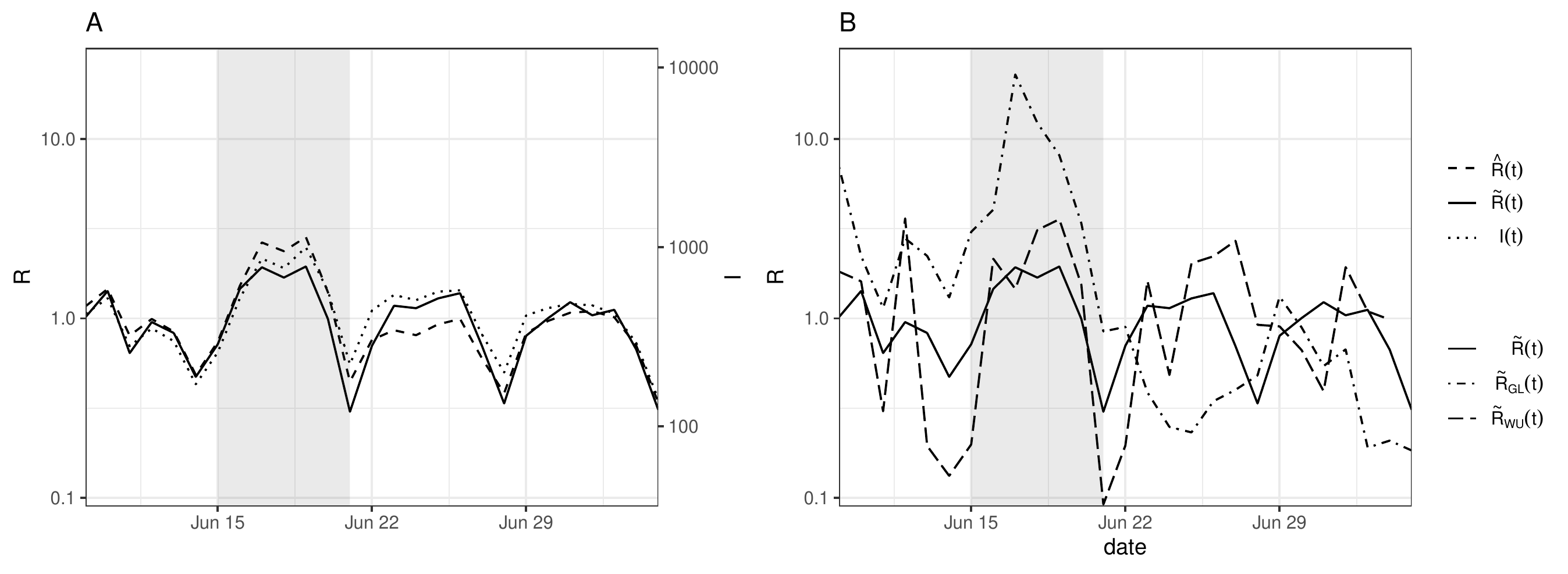}
    \caption{
        \textbf{A} shows reproduction number estimates and reported cases (\emph{dotted line}), both on a logarithmic scale, in Germany. On 17 June 2020 the first cases of a local outbreak were reported, causing a spike in the estimated reproduction numbers. Another consequence of this outbreak are lower estimates of the reproduction numbers (\emph{dashed line}) in the following weeks. Both phenomena are less pronounced for the estimate based on county level data (\emph{solid line}).
        \textbf{B} additionally shows county-level reproduction number estimates of Gütersloh county, $ \tilde R_{\mathrm{GL}} (t)$ (\emph{dot-dashed line}), and Wuppertal county, $ \tilde R_{\mathrm{WU}}(t) $  (\emph{double-dashed line}).
    }
    \label{fig:toennies}
\end{figure}
In the reproduction number estimation this causes undesirable artifacts: the nationwide spread of the epidemic is first overestimated due to the local outbreak while later the country-wide reproduction number will be underestimated since the denominator of $ \hat R (t) $ is too large due to the previous outbreak, for example Fig. \ref{fig:toennies} shows the effect of a huge influx of cases in June 2020 in Germany due to several smaller outbreaks, the biggest with $ 1413 $ cases occuring in a meat processing plant in Gütersloh county \cite{gunther_sars-cov-2_2020}. 

Small area estimation (SAE) is a branch of mathematical statistics providing tools suited for precisely this situation: data per region are scarce and may even be missing but there are many regions.
To make a virtue out of necessity, SAE models regional parameters as random variables, an approach we apply to county-level reproduction numbers.
Specifying the joint distribution of county-level reproduction numbers enables us to estimate a single set of parameters from which we can compute an estimated distribution of the reproduction number in each county.
This procedure can be viewed as empirical Bayes estimation.
We show that reproduction numbers obtained this way can be used to identify local outbreaks, handle low case numbers while agreeing with the country level estimates of the reproduction number \cite{hotz_monitoring_2020} in the absence of local outbreaks.

\section{Estimator}\label{sec:estimator}
A standard way of modeling the infection process is the renewal equation (\ref{eq:renewal}), cf. \cite{fraser_estimating_2007} for a detailed derivation. 
We present a straight forward generalization of this model to the regional level by using techniques from small-area estimation.
In small-area estimation it is common to model parameters on the regional level to vary randomly; in this spirit we model $ R_c(t) $, the regional reproduction number on day $ t $ in region $ c $, by a random variable. 

To account for cases that are imported and exported between regions, we assume that a fraction $ p_t $ of secondary cases are attributed to a region different than the corresponding primary case.
Let $ \Phi_c(t) = \sum_{\tau = 1}^{\infty} I_c(t - \tau)  w(\tau) $ be the expected number of active cases on day $t$ in county $ c $ given the past where $ I_c(t) $ denotes the incidences in that region on that day.
We then use the following renewal equation to describes the spread of the epidemic, relating the conditional distribution of $ I_c(t) $ to the expected number of active cases and the regional reproduction number $ R_c(t) $:
\begin{equation}
    \label{eq:renewal_pt}
    I_c(t) | R_c(t), I_c(t - 1), I_{c'}(t-1) \dots  \sim \mathrm{Pois} \big( R_c(t) \big( (1 - p_t) \Phi_c(t) + \textstyle\frac {p_t}{K -1} \sum\limits_{c' \neq c} \Phi_{c'}(t) \big)  \big)
\end{equation}
Here $ K $ denotes the total number of regions considered.
Note that we condition not only on past incidences $ I_c(t - \tau) $ in all counties but also on the random reproduction number $ R_c(t) $.

The interpretation of (\ref{eq:renewal_pt}) is straight-forward:
on day $ t $ there are $ I_c(t - \tau) $ individuals $ \tau $ days into their infection, thus $ R_c(t)w(\tau) I_c(t-\tau) $ is the expected amount of secondary infections caused by these individuals on day $ t $.
To account for the transfer of cases between counties, a fraction of $ p_t $ cases are counted towards the active cases in other regions and the wrongfully attributed cases are distributed equally among all other regions.
Summing over $ \tau $ yields the new infections $ I_c(t) $ which we assume to be Poisson distributed.

To infer $ R_c(t) $ from (\ref{eq:renewal_pt}) further assumptions about both the distribution of $ R_c(t) $ and the joint distribution of the pairs $ (I_c(t), R_c(t))$ for all regions $ c $ are necessary.
To this end we assume that the regional reproduction numbers on day $ t $ posses a common, known distribution and that the set of tuples $ (I_c(t), R_c(t)) $ is conditionally independent (given past incidences). 
More concretely, we assume the common distribution of the regional reproduction numbers $ R_c(t) $ to be a gamma distribution $ \mathrm{Gamma}(a_t, s_t) $ with \emph{shape} $ a_t $ and \emph{scale} $ s_t $ and density $ \frac {1}{s_t^{a_t} \Gamma(a_t)} x^{a_t-1} \exp \left(\frac {-x}{s_t}\right)$.

It is easy to see that the marginal distribution of $ I_c(t) $ (given the past incidences) in that region --- without conditioning on the reproduction number $ R_c(t) $ ---, is then a mixture of a gamma and a Poisson distribution, i.e. a negative binomial distribution whose parameters only depend on the parameters $ a_t, s_t, p_t $, past incidences $ I_c(t - \tau) $ and the generation time distribution $ w $.

As the conditional distribution of $ I_c(t) $ only depends on the unknown parameters, and, conditionally, the incidences of different regions are independent, we can apply maximum-likelihood estimation to obtain estimates $ \hat a_t $, $ \hat s_t $ and $ \hat p_t $ of the unknown parameters.

Also, the gamma distribution is conjugate prior to the Poisson distribution whence the conditional distribution of $ R_c(t) $ given past incidences $ I_c(t), I_c(t-1), \dots $ is again a gamma distribution whose shape and scale only depend on the unknown parameters and past incidences.
Thus one can use plug-in to estimate parameters of the posterior distribution such as $ \mathbf E \left(R_c(t)| I_c(t), \dots \right) $ and to derive prediction intervals.
Furthermore we naturally obtain a new estimator of the country-wide reproduction number, the estimated mean  $\tilde R(t) = \hat a_t \hat s_t $.

This approach could also be interpreted in the setting of empirical Bayes methods if one thinks of $\mathrm{Gamma}(a_t,s_t)$ as the prior distribution of $ R_c(t) $ and $ I_c(t) $ as the observations, with the prior parameters being estimated with tools from frequentist statistics. 

\section{Parameters, Data Sources and Implementation Details}
\label{sec:data}
The estimators consider assume the probability mass function $ w $ of the generation time to be known. 
As a precise model for the generation time is difficult to obtain we opt for a simple model: we assume the shape of $w$ to be trapezoidal with a mean of $ 5.6 $ days in accordance with the mean serial interval of 5.4 days found in \cite{zhang_meta-analysis_2020}, see \cite{hotz_monitoring_2020} for details.
In the same spirit we assume that the generation time distribution does not change over time.

To estimate the county-level reproduction numbers in Germany we use data provided by the Robert-Koch Institut \cite{robert_koch-institut_rki_2020}, as of 30 August 2021.
This dataset contains daily information on reported cases and deaths in Germany in addition to the county (Landkreis) where the case was reported to local health authorities.
There is a strong weekday effect present in both the case and death counts. 
This effect is most likely due to testing, evaluating tests and reporting occuring more frequently on workdays compared to weekends.
We do not account for this effect to direct the readers attention to the existence of such artifacts in the data and to avoid overconfidence in the resulting estimates --- these should be interpreted qualitatively not quantitatively.

Note that there is a delay between infection and reporting of cases so that estimates of reproduction numbers $ \hat R(t), \tilde R(t) $ ought to be backdated by about $ 7 $ days, see \cite{hotz_monitoring_2020} for details.



All computations, including simulations to validate the estimator, are conducted in \texttt{R} version 4.1.1 \cite{r_core_team_r_2021}. 
The calculation of maximum-likelihood estimates $ \hat a_t, \hat s_t, \hat p_t $ cannot be performed analytically, and is achieved using numerical optimization by the built-in function \texttt{optim}.

\section{Validation by Simulation}
\label{sec:validation}
To check how a mismatch between our model and reality might affect our estimator, we simulate a point process on the flat torus $ \mathbf T = \mathbf R^2 / (k \mathbf Z)^2 $, $ k \in \mathbf N$  where each of the $ k^2 $ unit squares corresponds to a county. 
We chose $ k = 20 $ to obtain $ k^2 = 400 $ counties, approximating the $ 401 $ counties in Germany. Time is chosen to be discrete and measured in days.
To simplify computation we simulate on $ \mathbf R^2 $ and quotient out $ (k \mathbf Z)^2 $ after the simulation has finished.

We initialize the simulation with $ 400 $ infected individuals that are placed uniformly on $ \mathbf T $, their infection age chosen again uniformly from the discrete support of the trapezoidal generation time distribution $ w $ (see Sect. \ref{sec:data}). 
At each time $ t $ every infected individual with infection age $ \tau $ in county $ c $ infects a random, $ \mathrm{Pois}(R_c(t) w(\tau)) $-distributed, number of new cases.

The position of the new cases is also random, and sampled from a bivariate normal distribution centered at the position of the primary case with covariance matrix $ \sigma^2 \mathbf I_2 $. 
We chose $ \sigma^2 $ such that approximately $ 20\% $ of secondary cases occur in counties different from their primary case, resulting in $ \sigma^2 \approx (0.14)^2 $.

These simulations introduce a mismatch between model (\ref{eq:renewal_pt}) and the generated incidence data. 
Firstly, exported cases are no longer distributed evenly over all counties, but rather depending on proximity. 
Secondly, we can choose the reproduction numbers to deviate from the assumed Gamma distribution.
To incorporate the introduction and partial lifting of non-pharmaceutical interventions we set $ R(t) $ to be $ 2.5 $ for $ 20 $ days, $ 0.7 $ for $ 40 $ days and $ 1.2 $ for another $ 40 $ days, simulating an outbreak over a total of $ 100 $ days.

\begin{figure}[t]
    \centering
    \includegraphics[width=\textwidth]{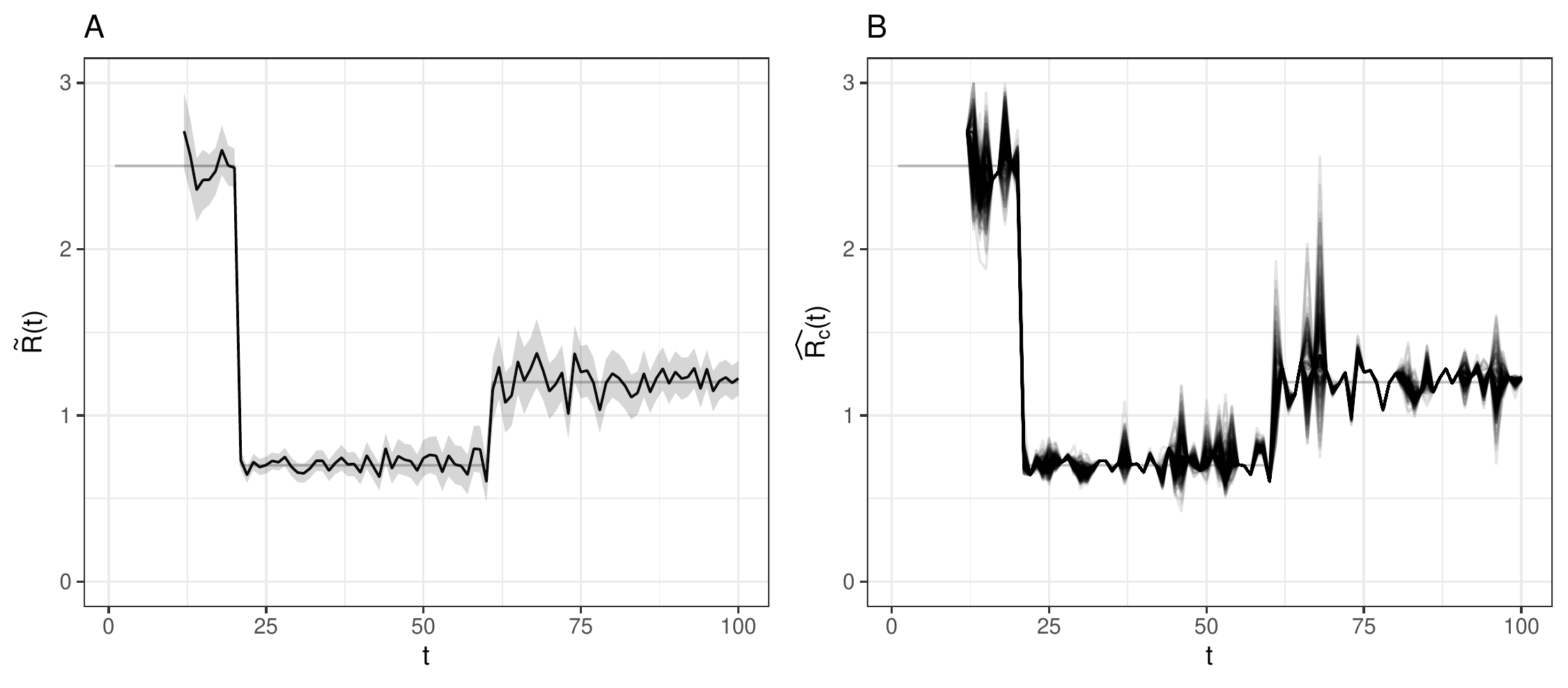}
    \caption{Results for one simulated outbreak. 
        \textbf{A} shows the estimates of the posterior mean $ \hat a_t \hat s_t $ in \emph{black} with corresponding confidence intervals indicated by \emph{grey ribbons}, the true $ R(t) $ is shown as a \emph{transparent grey line}. 
\textbf{B} shows the estimates of reproduction numbers on the county level.
}
    \label{fig:scenario_I}
\end{figure}

The daily reproduction number estimates based on the case data of this simulation as well as asymptotic $ 95\% $ confidence sets, based on the Fisher information, are shown in Fig. \ref{fig:scenario_I} \textbf{A}.
Despite the model mismatch the coverage of the confidence intervals is close to $ 95\% $ and also stays this way if we simulate this scenario multiple times (figures not shown).
Additionally the sharp changes in the reproduction number on days $ 21 $ and $ 61 $ are captured by our estimator as well.

We also show an estimate of $ \mathbf E \left( R_c(t) \middle| I_c(t), \dots \right) $, the county level reproduction numbers, for every county in Fig. \ref{fig:scenario_I} \textbf{B}. 
In this model the county level reproduction numbers have zero variance. 
This results in some estimates of the variance $ a_ts_t^2 $ to be very small, making all county level estimates similar at some time points. 
Increasing the regional variation by sampling reproduction numbers from a Gamma distribution did not produce such effects (figures not shown).


\section{Application to the COVID-19 Pandemic in Germany}
In Fig. \ref{fig:toennies} we depict our new estimator $ \tilde R(t) $ with $ \hat R(t) $ for Germany, with a special focus on the aforementioned outbreak in June 2020.
The weekly pattern in the estimates is due to the similar pattern in the incidence data; we decided against smoothing the estimates to highlight these complications with the data quality.
%
Note that in the week corresponding to the outbreak, $ \tilde R(t) $ is lower than the previous estimate.
Additionally, the downwards trend of $ \hat R (t) $ in the following weeks with estimates below $ 1 $ is no longer present, as the outbreak was a local one in few counties.
Except for the deviations mentioned above, $ \tilde R(t) $ resembles $ \hat R(t) $ remarkably well.
Around October 2020 a second wave of infections started to occur in Germany with rapidly rising case numbers across the country. 
Figure \ref{fig:toennies} shows that under these circumstances, i.e., high incidences in all regions, the country level estimates based on the small area estimation approach do not differ much from the estimates based on the country level.

\section{Discussion}\label{sec:discussion}
Of course our estimator rests on assumptions which ought to be discussed.
Modeling $ R_c(t) $ as random is a standard approach in small area estimation when dealing with few or even missing observations on a sub-national level; it is required to reduce the dimensionality of the parameter space.
For this, we critically assumed that on a fixed day $ t $ the regional reproduction numbers $ R_c(t) $ in different counties are independent and identically distributed according to a gamma distribution.
This is questionable as transmission dynamics vary with local social and economic factors.
For example one might expect that reproduction numbers are higher in urban counties than in rural counties with less population density.
Furthermore neighboring counties might exhibit spatial correlation.
Such socio-economic factors might be incorporated as for generalized linear mixed effects models although it is not obvious which factors to include and how to model their influence on the parameters $ a_t $, $ s_t $ and $ p_t $. 

Assuming a gamma distribution for the regional reproduction numbers $ R_c(t) $  is mathematically convenient as it is the conjugate prior distribution to the Poisson distribution, so using plug-in to obtain estimates for the posterior parameters is easy. 
In addition the log-likelihood of the posterior predictive distribution can be calculated analytically which makes estimation fast.
The price we pay for this distributional assumption is that the gamma distribution is a relatively light-tailed distribution prohibiting it from fully incorporating superspreading events such as the investigated outbreak. 
For this outbreak the country level estimates provided by $ \tilde R(t) $ are still elevated when compared to the previous and next week (see Fig. \ref{fig:toennies}), which might be an artifact of our choice of distribution as well as the small-area approach which biases estimates towards the country-wide mean.
Changing the marginal distribution of $ R_c(t) $ would lead to a computationally more involved estimation procedure requiring numerical integration.
The results in Sec. \ref{sec:validation}, however, show that our estimators are rather robust against slight misspecification in the prior distribution.


In addition to the mathematical assumptions discussed above we also made some more subtle epidemiological assumptions.
%
%
To account for infections across regions we introduced the parameter $ p_t $, the proportion of cases that were attributed to a different region than the one where infection occurred.
The addition of $ p_t $ is essential to the model when considering periods where incidence is low, e.g. during the summer in Germany. 
Without modeling cross-county infections, counties which have reached incidence $ 0 $ for a prolonged period of time would never record new cases, and observing new cases in such a county would lead to a breakdown of the estimator as the observed data would have likelihood $ 0 $.
We assumed that such transferred infections spread evenly among the other counties and that the this spread is the same for all counties, though the results of Sec. \ref{sec:validation} suggest robustness against such a model mismatch.
This could be improved by spatial models for the transfer of cases, e.g. based on mobility data.

We also assume the generation time distribution $ w $ to be constant over time and to be known.
The sensitivity of our new estimator to misspecification in the generation time could easily be studied by adapting the simulations from Sect. \ref{sec:validation} to include such a mismatch between simulation and estimation.
As this sensitivity is not the main concern of this paper, we omit such an analysis but refer the reader to \cite{hotz_monitoring_2020}.

We caution the reader to interpret the estimations and predictions proposed in this paper quantitatively due to the restrictions mentioned above as well as the quality of the available data.
Nevertheless we believe that the presented estimation procedure can be used to yield qualitative insight about the behavior of sub-national spread of an epidemic when case counts are low.
In such scenarios our estimator $ \tilde R(t) $ is a better representation of the country-level spread of the epidemic because it is less affected by local outbreaks.

\bibliographystyle{plain}
\bibliography{references.bib}
\end{document}